\documentstyle[12pt]{article}

\catcode`\@=11
\long\def\@makefntext#1{
\protect\noindent \hbox to 3.2pt {\hskip-.9pt
$^{{\ninerm\@thefnmark}}$\hfil}#1\hfill}                %CAN BE USED

\def\@makefnmark{\hbox to 0pt{$^{\@thefnmark}$\hss}}  %ORIGINAL

\def\ps@myheadings{\let\@mkboth\@gobbletwo
\def\@oddhead{\hbox{}
\rightmark\hfil\ninerm\thepage}
\def\@oddfoot{}\def\@evenhead{\ninerm\thepage\hfil
\leftmark\hbox{}}\def\@evenfoot{}
\def\sectionmark##1{}\def\subsectionmark##1{}}

\renewcommand{\thefootnote}{\fnsymbol{footnote}}

\newcounter{sectionc}\newcounter{subsectionc}\newcounter{subsubsectionc}
\renewcommand{\section}[1] {\vspace*{0.6cm}\addtocounter{sectionc}{1}
\setcounter{subsectionc}{0}\setcounter{subsubsectionc}{0}\noindent
        {\normalsize\bf\thesectionc. #1}\par\vspace*{0.4cm}}
\renewcommand{\subsection}[1] {\vspace*{0.6cm}\addtocounter{subsectionc}{1}
        \setcounter{subsubsectionc}{0}\noindent
        {\normalsize\it\thesectionc.\thesubsectionc. #1}\par\vspace*{0.4cm}}
\renewcommand{\subsubsection}[1]
{\vspace*{0.6cm}\addtocounter{subsubsectionc}{1}
        \noindent
{\normalsize\rm\thesectionc.\thesubsectionc.\thesubsubsectionc.
        #1}\par\vspace*{0.4cm}}

\newcounter{appendixc}
\newcounter{subappendixc}[appendixc]
\newcounter{subsubappendixc}[subappendixc]

\renewcommand{\appendix}[1] {\vspace*{0.6cm}
        \refstepcounter{appendixc}
        \setcounter{figure}{0}
        \setcounter{table}{0}
        \setcounter{equation}{0}
        \renewcommand{\thefigure}{\Alph{appendixc}.\arabic{figure}}
        \renewcommand{\thetable}{\Alph{appendixc}.\arabic{table}}
        \renewcommand{\theappendixc}{\Alph{appendixc}}
        \renewcommand{\theequation}{\Alph{appendixc}.\arabic{equation}}
        \noindent{\bf Appendix \theappendixc #1}\par\vspace*{0.4cm}}

\def\abstracts#1{{

\centering{\begin{minipage}{12.2truecm}\footnotesize\baselineskip=12pt\noindent
        \centerline{\footnotesize ABSTRACT}\vspace*{0.3cm}
        \parindent=0pt #1
        \end{minipage}}\par}}

\renewenvironment{thebibliography}[1]
        {\begin{list}{\arabic{enumi}.}
        {\usecounter{enumi}\setlength{\parsep}{0pt}
\setlength{\leftmargin 1.25cm}{\rightmargin 0pt}
         \setlength{\itemsep}{0pt} \settowidth
        {\labelwidth}{#1.}\sloppy}}{\end{list}}

\newcounter{itemlistc}
\newcounter{romanlistc}
\newcounter{alphlistc}
\newcounter{arabiclistc}

\newcommand{\fcaption}[1]{
        \refstepcounter{figure}
        \setbox\@tempboxa = \hbox{\footnotesize Fig.~\thefigure. #1}
        \ifdim \wd\@tempboxa > 6in
           {\begin{center}
        \parbox{6in}{\footnotesize\baselineskip=12pt Fig.~\thefigure. #1}
            \end{center}}
        \else
             {\begin{center}
             {\footnotesize Fig.~\thefigure. #1}
              \end{center}}
        \fi}

\newcommand{\tcaption}[1]{
        \refstepcounter{table}
        \setbox\@tempboxa = \hbox{\footnotesize Table~\thetable. #1}
        \ifdim \wd\@tempboxa > 6in
           {\begin{center}
        \parbox{6in}{\footnotesize\baselineskip=12pt Table~\thetable. #1}
            \end{center}}
        \else
             {\begin{center}
             {\footnotesize Table~\thetable. #1}
              \end{center}}
        \fi}

\def\@citex[#1]#2{\if@filesw\immediate\write\@auxout
        {\string\citation{#2}}\fi
\def\@citea{}\@cite{\@for\@citeb:=#2\do
        {\@citea\def\@citea{,}\@ifundefined
        {b@\@citeb}{{\bf ?}\@warning
        {Citation `\@citeb' on page \thepage \space undefined}}
        {\csname b@\@citeb\endcsname}}}{#1}}

\newif\if@cghi
\def\cite{\@cghitrue\@ifnextchar [{\@tempswatrue
        \@citex}{\@tempswafalse\@citex[]}}
\def\citelow{\@cghifalse\@ifnextchar [{\@tempswatrue
        \@citex}{\@tempswafalse\@citex[]}}
\def\@cite#1#2{{$\null^{#1}$\if@tempswa\typeout
        {IJCGA warning: optional citation argument
        ignored: `#2'} \fi}}

 1
 1
 1

\font\ninerm=cmr9

\input epsf
\epsfverbosetrue
\begin{document}

\def\detac{$\Delta\eta_{c}$}
\def\gapf{$f(\Delta\eta_c)$}

\begin{flushright}
McGill-95/31\\
hep-ex/9509004
\end{flushright}
\vspace*{0.9cm}

\centerline{\normalsize\bf HARD PHOTOPRODUCTION}
\centerline{\normalsize\bf BY COLOUR SINGLET EXCHANGE AT HERA}
%\centerline{\large\bf Colour Singlet Exchange}
%\centerline{\large\bf in Hard Photoproduction}
\baselineskip=28pt

%\vfill
\vspace*{0.6cm}
\centerline{\footnotesize ZEUS Collaboration}
\baselineskip=22pt
\centerline{\footnotesize Talk given by LAUREL SINCLAIR
\footnote{Talk presented at {\it Photon '95},
10th International Workshop on Photon-Photon Collisions,
Sheffield, U.K., April 1995.}}
\baselineskip=16pt
\centerline{\footnotesize\it Department of Physics, McGill University}
\baselineskip=12pt
\centerline{\footnotesize\it Montreal, Quebec, H3A 2T8, Canada}
\centerline{\footnotesize E-mail: sinclair@desy.de}
\vspace*{0.3cm}

%\vfill
\vspace*{0.9cm}
\abstracts{A search for photoproduction events which contain a rapidity gap
between the two highest transverse energy jets has been conducted at HERA
using the ZEUS detector.  The jets have transverse energies greater than
6~GeV, and are separated by pseudorapidity intervals of up to four units.
The fraction of events containing a gap is measured as a function of the
gap-width.  It is expected that this gap-fraction will fall exponentially
with the gap-width, until the dominant gap-production mechanism becomes
colour singlet exchange, at which point it will plateau.
An indication of a plateau in the measured gap-fraction has been found, at a
higher level than that expected from electroweak exchange.}

%\vspace*{0.6cm}
\normalsize\baselineskip=15pt
\setcounter{footnote}{0}
\renewcommand{\thefootnote}{\alph{footnote}}

\section{Introduction}

It was suggested some time ago that hard interactions
mediated by the exchange of a colour singlet propagator could give rise to
large regions of rapidity containing very few final-state
particles\cite{dokshitzer}.
That study emphasized the importance of
this ``rapidity gap'' as a signature for Higgs production via
$W^{+}W^{-}$~fusion.
Building on these ideas, Bjorken\cite{bjorken} proposed to
study the multiplicity distribution in pseudorapidity ($\eta$) and
azimuth ($\phi$) of
the final state particles in dijet events, and to
look for an absence of particles between the edges of the two jet cones.
He also pointed out that a significant background to such
Higgs searches could come from hard diffraction processes, and that
the study of these ``pomeron-exchange'' events was itself of great
importance.  Indeed, a recent
next-to-leading order calculation\cite{zeppenfeld} of the two gluon,
colour singlet exchange amplitude has shown that emitted soft gluons do
not resolve the internal colour structure of the ``Low-Nussinov''
pomeron.
This confirms the intuitive notion that pomeron-exchange events should
have a similar radiation pattern to electroweak exchange events, and
should therefore lead to the formation of rapidity gaps.

Note that the square of the momentum transfer ($t$) carried by the exchanged
colour singlet
object in these events is large, which should allow for the cross section
to be calculated within the framework of perturbative QCD.  This study is
thus complementary to the studies of diffractive hard scattering with a
``forward gap'' (i.e.~low~$t$), which are used to determine the structure
function of the pomeron\cite{forwardg}.

Both D0\cite{D0} and CDF\cite{CDF} have reported the results of searches
for dijet events containing a rapidity gap between the two jets in
 $p\bar{p}$  collisions.  In this paper, we report the results of such a search
in $\gamma p$ interactions obtained from $e^{+}p$ collisions.  Fig.~1(a) shows
an example of colour singlet exchange at HERA in which  a parton in the
photon scatters from a parton in the
proton, via $t$-channel exchange of a colour singlet object.  An example of
the more common, non-colour singlet exchange mechanism is shown in Fig.~1(b).

\vspace{0.4cm}
\begin{center}
\leavevmode
\epsfysize=100.pt
\epsfbox{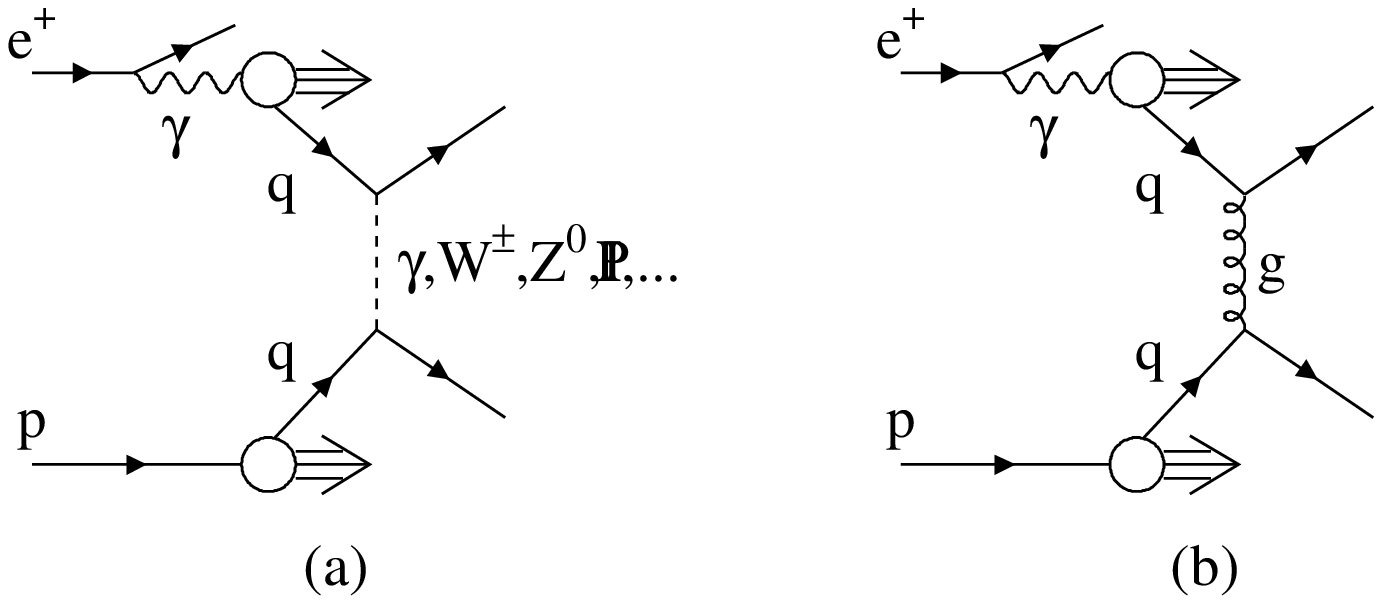}
\end{center}
\nopagebreak
\fcaption{Examples of (a) colour singlet and (b) non-colour singlet exchange
processes at HERA.}

\section{Rapidity Gap Signature}

The event topology for the processes of Fig.~1 is illustrated in Fig.~2(a).
There are two jets in the final state, which have a fixed cone radius
$R = \sqrt{\Delta\eta^2 + \Delta\phi^2}$.  \detac~ is defined as the
difference
in $\eta$ between the nearest tangents to the two jet cones.
For the colour singlet exchange processes of Fig.~1(a),
radiation is suppressed into the region of size \detac~ between the jet cones,
giving rise to the rapidity gap signature of these events.

\vspace{0.4cm}
\begin{center}
\leavevmode
\epsfysize=110.pt
\epsfbox{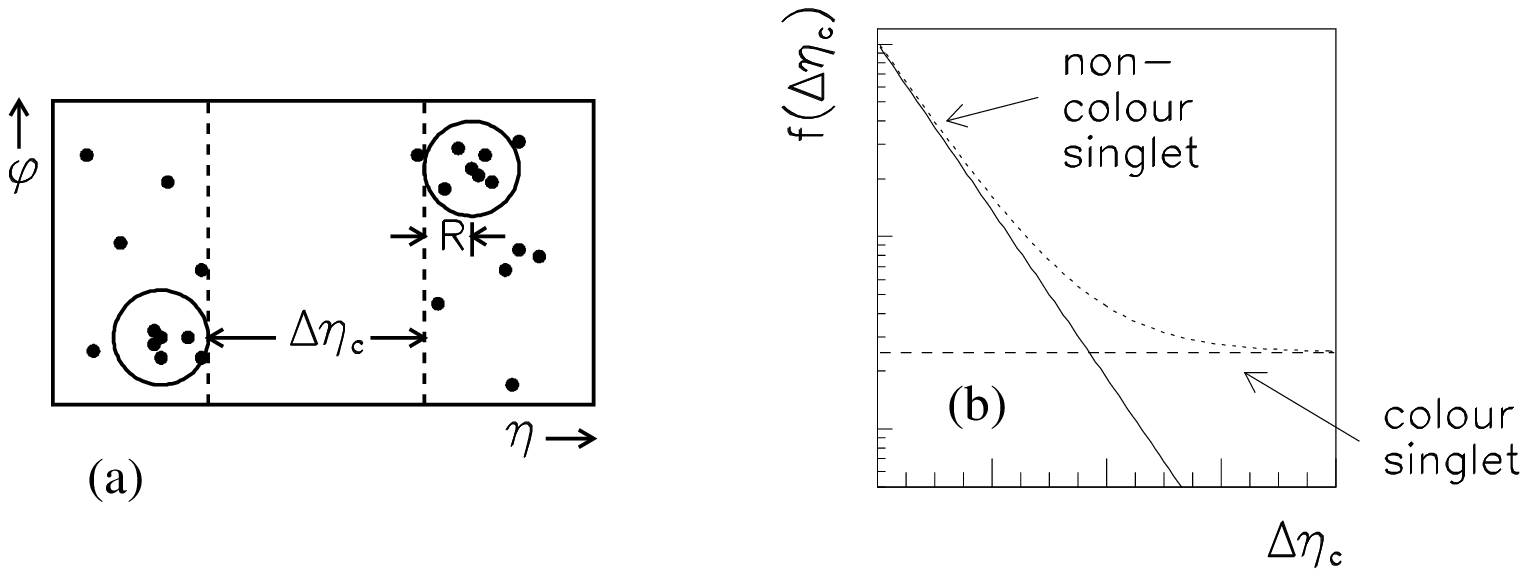}
\end{center}
\nopagebreak
\fcaption{Theoretical expectations for the event topology in colour singlet
events (a), and for the gap-fraction (b).  In (b) the non-colour singlet
component is shown as a solid line, the colour singlet component as a
dashed line, and the sum of the two as a dotted line.}

The quantity which we intend to measure is the fraction of
dijet events which contain a gap, as a function of the gap-width,
\begin{equation}
     f(\Delta\eta_c) \equiv \frac{d\sigma_{gap} / d\Delta\eta_c}
                                 {d\sigma / d\Delta\eta_c},
\end{equation}
where $d\sigma / d\Delta\eta_c$ is the dijet cross section as a function of
\detac, and $d\sigma_{gap} / d\Delta\eta_c$ is the cross section for dijet
events containing a rapidity gap between the two jets, as a function of \detac.
Note that $\sigma_{gap}$ accounts for the probability that a
gap between the two hadron jets has been filled in by secondary interactions
of the photon and proton spectator particles.  That is,
\begin{equation}
     \sigma_{gap} = \hat{\sigma}_{gap} \cdot {\cal P} (S)
\end{equation}
where $\hat{\sigma}_{gap}$ is the underlying cross section for the production
of
a gap between the two jets and ${\cal P}(S)$ is the probability that the
gap is not filled in by secondary interactions (the survival probability).
$\hat{\sigma}_{gap}$ may be further broken down as,
\begin{equation}
    \hat{\sigma}_{gap} = \hat{\sigma}_{gap}^{non-singlet} +
                                \hat{\sigma}_{gap}^{singlet},
\end{equation}
and these two contributions to the naive expectation for the behaviour of
\gapf~ are shown separately in Fig.~2(b).  It is expected that for low values
of \detac, multiplicity fluctuations in non-colour singlet exchange events
will give rise to a large fraction of gap events.  This component is
exponentially suppressed however, and at large \detac~ the dominant
contribution
to the gap-fraction should come from colour singlet exchange.

\section{Event Selection}

ZEUS has three levels of online triggering.  For this analysis we require
events to have some energy deposited in the calorimeter at the first trigger
stage.  At the second level we require high total transverse energy and at
the third level we are able to trigger on jets found using a fast cone
algorithm
in the calorimeter.  In addition, offline we require a good reconstructed
vertex
which  rejects proton beam-gas events and cosmic ray events.  Events with a
good scattered positron candidate in the calorimeter are rejected.  This
restricts the range of the photon virtuality to $P^{2} < 4$~GeV$^{2}$, and
results in a median $P^{2}$ of $\sim 10^{-3}$~GeV$^2$.  Charged current events
are rejected by requiring the calorimeter measurements of the missing
transverse
momentum, $\vec{p_T}$, and the total transverse energy, $E_T$, to satisfy
$\vec{p_T} / \sqrt{E_T} < 2$GeV$^{1/2}$.
We apply the cut $0.15 \leq y < 0.7$ on the fraction of the
positron's momentum which is carried by the photon, where $y$ is reconstructed
using the Jacquet-Blondel method.  Finally a cone algorithm is applied to
the calorimeter
cells to find jets with a cone radius
$R = \sqrt{\Delta\eta^2_{cell} + \Delta\phi^2_{cell}} < 1$.  In order to have
high acceptance for dijet events with true $E_T^{jet} > 6$~GeV, we require the
events to have at least two calorimeter jets with transverse
energy $E_T^{jet} > 5$~GeV.  The jets must not be too close to the forward
proton direction, $\eta^{jet} < 2.5$.  The two highest transverse energy
jets are used in the calculation of \detac, and \detac~ must be greater
than 0 (i.e. the jet cones may not be overlapping in $\eta$).

In addition we group adjacent calorimeter cells which have a common energy
maximum
into ``islands''.  The multiplicity of particles between the jet cones is then
estimated by counting the number of islands $n$ which have transverse momentum
above 200~MeV and which lie between the jet cones in $\eta$.  Gap events have
$n = 0$.  Defining a momentum threshold makes the definition of a gap
insensitive
to noisy calorimeter cells, and also has theoretical advantages\cite{pumplin}.

{}From a sample of 2.4~pb$^{-1}$ of $e^{+}p$ collisions delivered by HERA in
1994,
$N = 17518$ dijet events were selected in this manner, of which
$N_{gap} = 5726$
have $n = 0$.
The non-$e^{+}p$ collision background has been estimated using the number of
events associated with unpaired bunch crossings.  We estimate the proton and
positron beam-gas contaminations to be about 0.1\%.  Cosmic ray contamination
is negligible.  The gap events which have $\Delta\eta_c > 1.5$ were also
scanned
visually to search for contamination from events where the energy deposits of
the
scattered positron or a prompt photon mimic a jet.  No such events were found.

\section{Acceptance and Resolution}

In order to study the acceptance of the ZEUS detector and the resolution
which it provides of the kinematic variables, we have generated a large
Monte Carlo sample.  The generator which we have used is PYTHIA\cite{pythia}
with the minimum $p_T$ of the hard scatter set to 2.5~GeV.  Resolved and
direct photon interactions are generated separately and are mixed according
to the cross sections determined by PYTHIA.  We have used the GRV parton
distribution for the photon and MRSA for the proton.  Two PYTHIA samples
will be treated separately in the following, and labelled ``reconstructed''
and ``hadronic''.

To obtain the reconstructed sample of PYTHIA events we have passed the
generated
final-state particles  through a detailed simulation of the ZEUS detector,
keeping only those events which pass all stages of data selection.
For this sample we use the simulated energy deposits in the calorimeter cells
as input to the jet-finding algorithm to determine $E_T^{jet}$,$\eta^{jet}$
and \detac.  The multiplicity $n$ is estimated using the islands of calorimeter
cells.
The selection criteria then correspond exactly to those applied to the data,
as outlined in Sec.~3.

The hadronic sample consists of the generated particle kinematics.  No detector
simulation is applied to these events.
The restriction $0.2 \leq y < 0.8$ is made using the true fraction of the
positron's momentum which is carried by the photon.
The jet finding algorithm is applied to the final state particles from the
event in order to determine $E_T^{jet}$,$\eta^{jet}$ and \detac.  Then it
is required that $E_T^{jet} > 6$ GeV and $\eta^{jet} < 2.5$ and that
$\Delta\eta_c > 0$ for the two highest transverse energy jets.
The higher $E_T^{jet}$
threshold for the hadronic sample reflects the experimental shift and
resolution of this quantity of about $-13\%$ and $13\%$ respectively.  The
angular resolution is about $9\%$ in $\eta^{jet}$ with negligible shift.
These values are consistent with those obtained during extensive studies
of the ZEUS 1993 hard photoproduction sample\cite{jonb}.  The multiplicity
is of course the number $n$ of final-state particles having transverse
energy $> 200$~MeV and lying between the jet cones. Gap events have $n = 0$.

The selection applied to the hadronic sample defines the true kinematic
range of interest in this study.  Any deviation of the reconstructed
PYTHIA distributions from the associated hadronic distributions is an
artificial detector effect.

\vspace{0.4cm}
\epsfxsize=15.2cm
\epsfbox{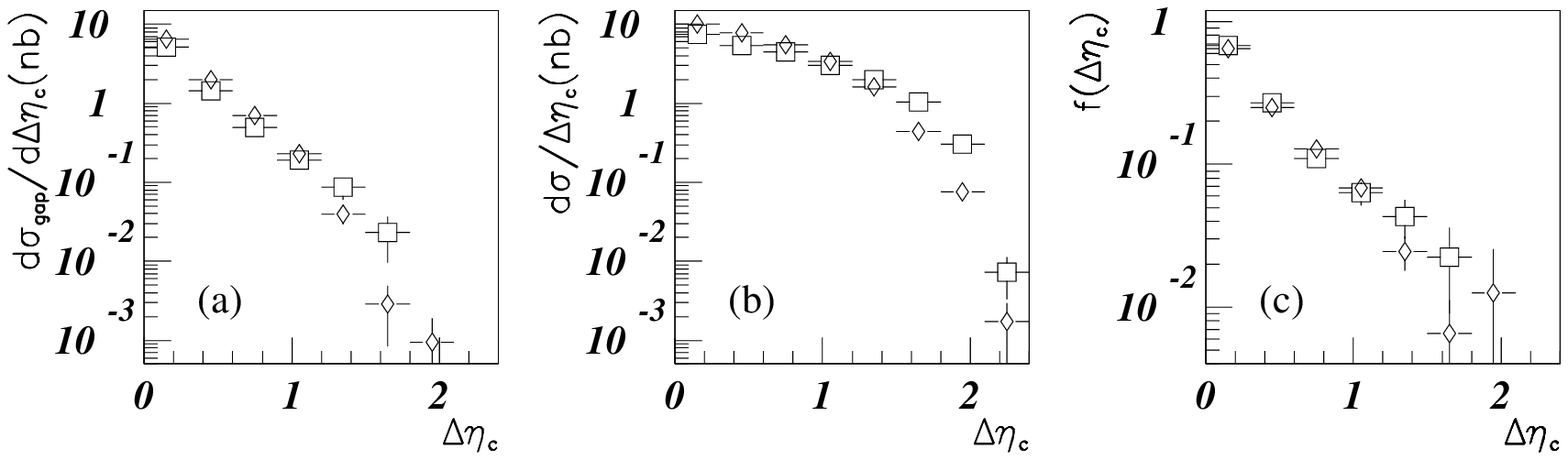}
\vspace*{0.2cm}
\nopagebreak
\fcaption{The distribution of the PYTHIA events with respect
to \detac.  Squares represent the ``reconstructed'' quantities and diamonds
the ``hadronic'' quantities (see text).  (a) shows the cross section for
producing
a gap, (b) is the inclusive cross section and (c) shows the gap-fraction.}

The distribution in \detac~ of the reconstructed and hadronic samples is
shown in Fig.~3.  The cross sections (according to PYTHIA) for producing
gap events are shown in Fig.~3(a).  Fig.~3(b) shows the inclusive dijet
differential
cross sections and Fig.~3(c) shows the gap-fraction (the ratio of Fig.~3(a) and
Fig.~3(b)).  It is clear from Fig.~3(a) and Fig.~3(b) that there are
significant
detector effects in the distribution of \detac.  However these effects
cancel when we take the ratio
of the gap cross section to the inclusive cross section as shown in Fig.~3(c).
This cancellation of acceptance corrections was expected,
as the selection criteria were designed to be sensitive only to the energies
and
angles of the jets, and not to the particle multiplicity between them.

\section{Results}

The results for the data are shown in Fig.~4.  These distributions are not
corrected
for detector effects.  The errors shown are statistical only.  The
distribution of
$N_{gap}(\Delta\eta_c)$ falls rapidly with \detac~ (Fig.~4(a)); however the
inclusive number of events $N(\Delta\eta_c)$ is also decreasing with \detac.
It is the number of events which contain a gap, $N_{gap}(\Delta\eta_c)$,
normalized to the total number of events, $N(\Delta\eta_c)$, which is
interesting.
This is the measured gap-fraction, \gapf, shown in Fig.~4(c).
The gap-fraction falls rapidly with \detac~ out to $\Delta\eta_c \sim 1.2$.
For larger values of \detac~ the gap-fraction does not appear to decrease as a
function of \detac.  It appears to exhibit a plateau, at a level of roughly
$4 \cdot 10^{-2}$.

\vspace{0.4cm}
\epsfxsize=15.2cm
\epsfbox{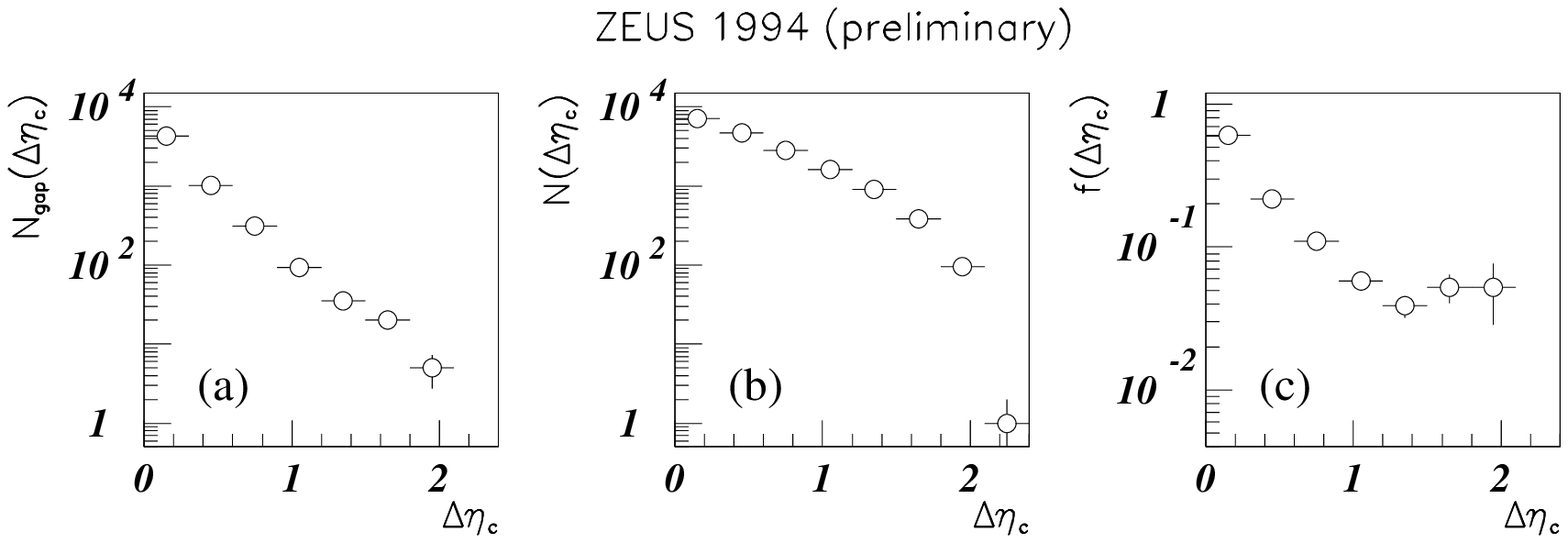}
\vspace*{0.2cm}
\nopagebreak
\fcaption{The distribution of the data with respect
to \detac.  In (a) is the number of events with a gap, and in (b) is the
total number of events.  (c) shows the gap-fraction.}

\section{Conclusions}

The conclusion of this study is that we have seen an indication of a
plateau of magnitude about $4 \cdot 10^{-2}$ in the behaviour of the measured
gap-fraction.  Monte Carlo studies indicate that the corrections
for detector effects will be quite small.  We are therefore encouraged to
further pursue the study of rapidity gaps between jets at HERA.

It is worth commenting on the significance of the
distribution we are measuring.  If the plateau is interpreted as evidence
for the exchange of a colour singlet particle, then one
can already make some restrictions on the identity of that particle.  The
gap-fraction from electroweak exchange is of order\cite{bjorken}
$\alpha^2/\alpha_s^2 \sim 0.001$.  The exchanged particle
could therefore not be an electroweak boson.  More
reasonable is an estimate\cite{bjorken} based on two-gluon exchange in
a colour singlet state,
$ \hat{\sigma}_{gap} / \sigma \sim 0.1$.  Such a model could account for our
data if the
survival probability in photoproduction reactions at HERA were roughly 0.5.

\end{document}